\documentclass[preprint,aps,amsmath,nofootinbib,english,11pt]{revtex4}
\usepackage{graphics,color,array,dcolumn}
\usepackage{graphicx}
\usepackage{amssymb}
\usepackage{xspace}
\usepackage{color}

\def\g#1{h^{(#1)}}

\makeatother
\begin{document}

\title{One loop beta functions and fixed points in Higher Derivative Sigma Models}
\bigskip
\author{Roberto Percacci\footnote{\it on leave from SISSA, via Beirut 4, I-34014 Trieste, Italy. Supported in 
part by INFN, Sezione di Trieste, Italy}}
\email{rpercacci@perimeterinstitute.ca}
\affiliation{Perimeter Institute for Theoretical Physics, 31 Caroline St. North, Waterloo, Ontario N2J 2Y5, Canada}
\author{Omar Zanusso}
\email{zanusso@sissa.it}
\affiliation{SISSA, via Beirut 4, I-34014 Trieste, Italy, and INFN, Sezione di Trieste, Italy}
\pacs{}
\medskip
\begin{abstract}
We calculate the one loop beta functions of nonlinear sigma models in four dimensions containing general
two and four derivative terms.
In the $O(N)$ model there are four such terms and nontrivial fixed points exist for all $N\geq4$.
In the chiral $SU(N)$ models there are in general six couplings, but only five for $N=3$
and four for $N=2$; we find fixed points only for $N=2,3$.
In the approximation considered, the four derivative couplings are asymptotically free
but the coupling in the two derivative term has a nonzero limit.
These results support the hypothesis that certain sigma models may be asymptotically safe.
\end{abstract} 

\maketitle

\section{Introduction}

In the study of quantum gravity one encounters many technical complications,
and it is often desirable to test one's ideas and tools in a simpler setting.
The NonLinear Sigma Models (NLSMs) have striking similarities to gravity:
they are nonpolynomially interacting theories, and from the point of view
of power counting, they have exactly the same structure as gravity.
On the other hand, they lack the complications due to gauge invariance.
They are therefore a good theoretical laboratory where one can study various
technical aspects of the renormalization of gravity without having to
consider the complications due to gauge fixing, and with the certainty
that one's results are not gauge artifacts.
Recent work on the beta functions of gravity suggests that there
might exist a nontrivial fixed point with finitely many UV attractive directions,
making this theory ``asymptotically safe''.
This means that if one considers all the terms in the derivative expansion
of the effective action, the corresponding (renormalized) couplings would
all run towards a Fixed Point (FP) in the UV limit, and that only finitely many 
combinations of couplings would be relevant (attracted to the FP in the UV).
Then, the requirement of tending to the FP in the UV would constrain the theory
to lie on a finite dimensional surface, and the theory would then be predictive.
See the original work \cite{weinberg} for the definition of asymptotic safety,
\cite{reviews} for reviews and \cite{cpr,bms} for more recent results.
Understanding the UV behaviour of the NLSM may shed some light on 
the analogous issue for gravity.

Aside from this,
the NLSMs also play an important role in particle physics phenomenology:
they are used as low energy effective field theories both for strong
and weak interactions. In the former case the scalar fields are identified with the
light mesons \cite{gl}, in the latter with the three Goldstone degrees of freedom
of the complex Higgs doublet \cite{ab}.
These effective field theories are usually thought to break down at some cutoff scale,
of the order of the GeV in the strong case and of the TeV in the weak case.
It is an interesting question in itself, and one that may have some relevance 
also for particle physics, whether some of these NLSM's
might actually be asymptotically safe.
Old work on the epsilon expansion and $1/N$ expansion suggests that a fixed
point with the right properties may exist \cite{polyakov,zinnjustin,bardeen,arefeva}. 
More recently, the beta functions of the NLSM were recalculated using
a two derivative truncation of an exact RG equation,
and it was found in the case of the $O(N)$ models
that they have a nontrivial UV FP \cite{codello2}.
In the present work we begin addressing the issue of asymptotic safety in the NLSM
taking into account also four derivative interactions.
The beta functions of four derivative NLSM were considered before in 
\cite{hasenfratz} and \cite{bk}.
The former reference uses a formalism that applies only to group-valued models;
the latter uses dimensional regularization and therefore cannot properly compute the
running of the two derivative terms, which is necessary to establish asymptotic
safety. In this paper we extend and partly correct the results of these earlier
works.

This paper is organized as follows: in section 2 we discuss
the models and the techniques we use; in section 3 we evaluate the beta functions,
first for arbitrary manifolds and then for the $O(N)$ models and the chiral models; 
in section 4 we list their fixed points; 
in section 5 we close with a discussion.
The comparison with \cite{hasenfratz} is given in appendix A.

\section{The theory}

\subsection{Geometry and action}

In general the NLSM is a field theory whose configurations are maps
from $\varphi:X\to Y$, where $X$ is a $d$-dimensional manifold
interpreted as spacetime and $Y$ is some $n$-dimensional internal manifold.
We will always take $X$ to be four dimensional and to have a fixed flat Euclidean metric,
and we will call $h$ a riemannian metric on $Y$.
Given a map $\varphi$, one calls ``vectorfield along $\varphi$''
a rule that assigns to each point $x$ of $X$ a vector tangent to $Y$ 
at $\varphi(x)$.
\footnote{The vectorfields along $\varphi$ should be thought of, in geometrical terms,
as sections of the pullback bundle $\varphi^*TY$.}
For example, given a fixed vector $v$ tangent to $X$ at $x$,
the image of $v$ under the tangent map $T\varphi$ is a vectorfield along $\varphi$.
Its components are $v^\mu\partial_\mu\varphi^\alpha$.
Thus we can view the matrix $\partial_\mu\varphi^\alpha$ as the components of four
vectorfields along $\varphi$.

The Levi-Civita connection of the metric $h$ in $TY$  
can be used to define the covariant derivative of vectorfields
along $\varphi$.
Let $\mit\Gamma_\alpha{}^\beta{}_\gamma$ be the Christoffel symbols of $h$ and 
$R_{\alpha\beta}{}^\gamma{}_\delta=\partial_\alpha{\mit\Gamma}_\beta{}^\gamma{}_\delta
-\partial_\beta{\mit\Gamma}_\alpha{}^\gamma{}_\delta
+{\mit\Gamma}_\alpha{}^\gamma{}_\epsilon{\mit\Gamma}_\beta{}^\epsilon{}_\delta
-{\mit\Gamma}_\beta{}^\gamma{}_\epsilon{\mit\Gamma}_\alpha{}^\epsilon{}_\delta$ 
its Riemann tensor.
The covariant derivative of a vectorfields along $\varphi$ is
\begin{equation}
\label{covariant}
\nabla_\mu\xi^\alpha=
\partial_\mu\xi^\alpha+\partial_\mu\varphi^\gamma{\mit\Gamma}_\gamma{}^\alpha{}_\beta\xi^\beta
\end{equation}

A diffeomorphism $f$ of $Y$ can be represented in coordinates by $y'=f(y)$.
It maps vectorfields along $\varphi$ to vectorfields along $\varphi'=f\circ\varphi$.
One can check explicitly using the transformation properties
\begin{equation}
\xi'^\alpha={\partial\varphi^{\prime\alpha}\over\partial\varphi^\beta}\xi^\beta\ ;\qquad
{\mit\Gamma}'{}_\gamma{}^\alpha{}_\beta
={\partial \varphi^\eta\over\partial \varphi^{\prime\gamma}}
{\partial \varphi^{\prime\alpha}\over\partial \varphi^\delta}
{\partial \varphi^\epsilon\over\partial \varphi^{\prime\beta}}
{\mit\Gamma}_\eta{}^\delta{}_\epsilon
+{\partial \varphi^{\prime\alpha}\over\partial \varphi^\delta}
{\partial^2 \varphi^\delta\over\partial \varphi^{\prime\gamma}\partial \varphi^{\prime\beta}}
\end{equation}
that the covariant derivative transforms in the same way as $\xi$ under diffeomorphisms of $Y$.

We also note for future reference that the curvature of the pullback connection 
is the pullback of the curvature of the Levi-Civita connection:
\begin{equation}
\label{curvature}
[\nabla_\mu,\nabla_\nu]\xi^\gamma
\equiv\Omega_{\mu\nu}{}^\gamma{}_\delta\xi^\delta
=\partial_\mu\varphi^\alpha\partial_\nu\varphi^\beta
R_{\alpha\beta}{}^\gamma{}_\delta \xi^\delta
\end{equation}

We can now discuss the dynamics of the NLSM.
Since the ordinary derivatives of $\varphi^\alpha$ are the components of vectorfields
along $\varphi$, the second covariant derivatives of the scalars are given by
\begin{equation}
\nabla_\mu\partial_\nu\varphi^\alpha=
\partial_\mu\partial_\nu\varphi^\alpha
+\partial_\mu\varphi^\beta{\mit\Gamma}_\beta{}^\alpha{}_\gamma
\partial_\nu\varphi^\gamma\ .
\end{equation}
Note that due to the symmetry of the Christoffel symbols 
$\nabla_\mu\partial_\nu\varphi^\alpha=\nabla_\nu\partial_\mu\varphi^\alpha$.
We also define $\Box\varphi^\alpha=D^\mu\partial_\mu\varphi^\alpha$.
After these preliminaries, the most general Lorentz-- and parity--invariant
NLSM with up to four derivatives has an action of the form:
\begin{eqnarray}
\label{action}
&\frac{1}{2}\int d^4x\,\Bigl[&\!\!\!
\partial_{\mu}\varphi^\alpha\partial^\mu\varphi^\beta \g2_{\alpha\beta}(\varphi)
+\Box\varphi^\alpha \Box\varphi^\beta \g4_{\alpha\beta}(\varphi)
\nonumber
\\
&&+\nabla_\mu\partial_{\nu}\varphi^\alpha\partial^\mu\varphi^\beta\partial^\nu\varphi^\gamma
A_{\alpha\beta\gamma}(\varphi)
+\partial_{\mu}\varphi^\alpha\partial^\mu\varphi^\beta 
\partial_{\nu}\varphi^\gamma\partial^\nu\varphi^\delta T_{\alpha\beta\gamma\delta}(\varphi)
\Bigr]
\ .
\end{eqnarray}
Here we defined parity to correspond to the reflection 
$\varphi^\alpha(x_1,x_2,x_3,x_4)\mapsto \varphi^\alpha(-x_1,x_2,x_3,x_4)$.
This is the only parity operation one can define in full generality.
We will discuss below other ``parities'' that can be defined
on special manifolds.
At the classical level, $\g2$, $\g4$ $A$, and $T$ are fixed tensorfields on $Y$.
They represent, in general, an infinite number of interaction terms.
In the quantum theory these tensors will be subject to RG flow.
The tensors $\g2$, $\g4$ are assumed to be positive definite metrics.
In the present work we will always use $\g4$ to raise and lower indices,
while $\g2$ is treated as any tensor. Of course nothing ultimately can depend
on this convention.
The tensor $A$ can be assumed to be totally symmetric without loss of generality.
The tensor $T$ must have the following symmetry properties:
$$
T_{\alpha\beta\gamma\delta}=
T_{\beta\alpha\gamma\delta}=
T_{\alpha\beta\delta\gamma}=
T_{\gamma\delta\alpha\beta}\,.
$$

In \eqref{action} we have not considered (parity violating) terms that involve 
the $\epsilon$ tensor, of the form
\begin{equation}
c\int d^4x\,\epsilon^{\mu\nu\rho\sigma}
\partial_\mu\varphi^\alpha\partial_\nu\varphi^\beta 
\partial_\rho\varphi^\gamma\partial_\sigma\varphi^\delta 
B_{\alpha\beta\gamma\delta}(\varphi)\,,
\end{equation}
where $B$ is some four-form on $Y$. These could be called ``Wess-Zumino-Witten terms''
in a generalized sense. A proper Wess-Zumino-Witten term is one for which the
four form $B$ is not defined everywhere on $Y$, but the five-form $H=dB$ is.
Then $H$ defines a nontrivial fifth-cohomology class and the coefficient $c$
has to obey a quantization condition.
The original Wess-Zumino term corresponds to the case $Y=SU(N)$
and $H=\mathrm{tr}(g^{-1}dg)^5$.
We will briefly return to these terms in the discussion.

We observe that since the field $\varphi$ appears nonpolynomially
in the action, it must be dimensionless. 
Then, $\g2$ must have dimension of mass squared, whereas
the other tensors are dimensionless.
Later on we will find it convenient to split off a dimensionful coupling from
the dimensionful tensors, so that all the tensors are dimensionless.

We will be especially interested in cases in which the theory has some global
symmetries. Let $\Phi$ be a diffeomorphism of $Y$ that leaves the tensors
$\g2$, $\g4$, $A$, $T$ invariant, for example
$$
T_{\alpha\beta\gamma\delta}(y)=
\frac{\partial\Phi^{\alpha'}}{\partial y^\alpha}
\frac{\partial\Phi^{\beta'}}{\partial y^\beta}
\frac{\partial\Phi^{\gamma'}}{\partial y^\gamma}
\frac{\partial\Phi^{\delta'}}{\partial y^\delta}
T_{\alpha'\beta'\gamma'\delta'}(\Phi(y))\ .
$$
In particular, $\Phi$ is an isometry of $\g4$.
Then the action is invariant under the transformation $\varphi\mapsto \Phi\circ\varphi$.
Such global symmetries may be discrete, or they may form a continuous group $G$.
In the latter case there exist vector fields $K_a$ on $Y$ (with $a=1\ldots\mathrm{dim}G$)
whose Lie brackets form an algebra isomorphic to the Lie algebra of $G$,
and such that $\g2$, $\g4$, $A$, $T$ are invariant under $G$:
\begin{equation}
\nonumber
{\cal L}_{K_a}\g2=0\ ;\quad
{\cal L}_{K_a}\g4=0\ ;\quad
{\cal L}_{K_a}A=0\ ;\quad
{\cal L}_{K_a}T=0\ .\quad
\end{equation}
In particular, $K_a$ are Killing vectors for the metric $\g4$:
$\nabla_\alpha K_{a\beta}+\nabla_\beta K_{a\alpha}=0$.
Then, the action \eqref{action} is invariant under the
infinitesimal transformation $\delta_\epsilon\varphi^\alpha=\epsilon^a K_a^\alpha(\varphi)$.

Discrete isometries may appear in the definition of parity or time reversal.
In linear scalar theories one can define the operation $\phi\mapsto -\phi$.
For example the pions transform as 
$(P\pi)^a(x_1,x_2,x_3,x_4)=-\pi^a(-x_1,x_2,x_3,x_4)$ under parity.
In a general NLSM the transformation
$\varphi^\alpha\mapsto -\varphi^\alpha$ has no intrinsic meaning.
However, suppose that every point $y\in Y$ is the fixed point 
of an involutive isometry $\Phi_y$.
Such a manifold is said to be a symmetric space \cite{helgason}.
We can then define a new parity operation, let us call it ``Parity''
with capital P, by 
$(P\varphi)^\alpha(x_1,x_2,x_3,x_4)=\Phi_0\circ\varphi(-x_1,x_2,x_3,x_4)$,
where $\Phi_0$ is the involutive isometry of the vacuum.
The transformation properties of the action under this new definition
of parity are different than under the previous definition.
In particular, if 
$A_{\alpha\beta\gamma}(\Phi_0(y))=A_{\alpha\beta\gamma}(y)$,
then the $A$-term will not be Parity--invariant.
On the other hand if 
$B_{\alpha\beta\gamma\delta}(\Phi_0(y))=-B_{\alpha\beta\gamma\delta}(y)$,
then the Wess-Zumino-Witten term is Parity--invariant
\cite{witten}.

\subsection{Background field expansion}

We use the background field techniques developed in \cite{honerkamp,afm,bb,hps}.
We review here some of the main points.
Having chosen a (not necessarily constant) background $\bar\varphi$,
any other field $\varphi$ in an open neighborhood of $\bar\varphi$ can be written
$\varphi^\alpha=\bar\varphi^\alpha+\eta^\alpha$.
In principle one could work with the quantum fields $\eta^\alpha$,
but this is not convenient because, as differences of coordinates, they do not have
nice transformation properties.
It is therefore convenient to proceed as follows.
For each $x$ one can find a unique vector $\xi(x)$ tangent to $\bar\varphi(x)$
such that $\varphi(x)$ is the point on the geodesic passing through
$\bar\varphi(x)$ and tangent to $\xi(x)$, the distance between $\varphi(x)$ and $\bar\varphi(x)$
being equal to $|\xi(x)|$.
We can thus write $\varphi(x)=Exp_{\bar\varphi(x)}\xi(x)$, where $Exp$ is the exponential map.
The field, $\xi^\alpha(x)$ is a vectorfield along $\bar\varphi$,
and its covariant derivative is defined as in \eqref{covariant}.

In principle, then, the action $\varphi$ can be rewritten as $S(\varphi)=\bar S(\bar\varphi,\xi)$.
In practice one can compute the first few terms in an expansion 
$
\bar S(\bar\varphi,\xi)=\bar S^{(0)}(\bar\varphi,\xi)+\bar S^{(1)}(\bar\varphi,\xi)
+\bar S^{(2)}(\bar\varphi,\xi)+\ldots\ ,
$
where $\bar S^{(n)}$ contains $n$ powers of $\xi$.
The first term is clearly $\bar S^{(0)}(\bar\varphi,\xi)=\bar S(\bar\varphi,0)=S(\bar\varphi)$.
To compute the next terms we use the following formulae (whose derivation can be found in \cite{honerkamp}):
\begin{eqnarray}
\partial_\mu\varphi^\alpha&=&
\partial_\mu\bar\varphi^\alpha
+\bar\nabla_\mu\xi^\alpha
-\frac{1}{3}\partial_\mu\bar\varphi^\gamma\bar R_{\gamma\epsilon}{}^\alpha{}_\eta\xi^\epsilon\xi^\eta+\ldots
\nonumber\\
t_{\alpha\beta\ldots}(\varphi)&=&
t_{\alpha\beta\ldots}(\bar\varphi)
+\xi^\epsilon \bar\nabla_\epsilon t_{\alpha\beta\ldots}(\bar\varphi)
+\frac{1}{2}\xi^\epsilon\xi^\eta\bar\nabla_\epsilon\bar\nabla_\eta t_{\alpha\beta\ldots}(\bar\varphi)
-\frac{1}{6}\xi^\epsilon\xi^\eta \bar R^\gamma{}_{\epsilon\alpha\eta}t_{\gamma\beta\ldots}
-\frac{1}{6}\xi^\epsilon\xi^\eta \bar R^\gamma{}_{\epsilon\beta\eta}t_{\alpha\gamma\ldots}
+\ldots\nonumber
\end{eqnarray}
A bar over the derivatives and the curvatures indicates that they have to be computed
with the background field $\bar\varphi$.
In particular for the metric $g$ we have
$$
g_{\alpha\beta}(\varphi)=
g_{\alpha\beta}(\bar\varphi)
-\frac{1}{3}\bar R_{\alpha\epsilon\beta\eta}\xi^\epsilon\xi^\eta +\ldots
$$
Inserting in \eqref{action}, with $A=0$, and keeping terms of second order in $\xi$ we obtain
\begin{eqnarray}
\label{secondvariation}
&&\!\!\!\!\!\!\! \frac{1}{2}\int\! d^4x \Bigl[
\g2_{\alpha\beta} \nabla_\mu\xi^\alpha \nabla^\mu\xi^\beta
-\xi^\alpha\xi^\beta R_{\alpha\gamma\beta}{}^\epsilon
\g2_{\epsilon\delta}\partial_\mu\varphi^\gamma\partial_\mu\varphi^\delta
+2 \xi^\alpha\nabla_\mu\xi^\beta \nabla_\alpha\g2_{\beta\gamma} \partial_\mu\varphi^\gamma 
+\frac{1}{2}\xi^\alpha\xi^\beta \nabla_\alpha \nabla_\beta\g2_{\gamma\delta} 
\partial_\mu\varphi^\gamma\partial_\mu\varphi^\delta
\nonumber\\
&&
+\g4_{\alpha\beta}\Box\xi^\alpha\Box\xi^\beta
+2\xi^\alpha\Box\xi^\beta R_{\alpha\gamma\beta\delta}
\partial_\mu\varphi^\gamma\partial^\mu\varphi^\delta
-4\xi^\alpha\nabla_\mu\xi^\beta R_{\alpha\gamma\beta\delta}
\partial^\mu\varphi^\gamma \Box\varphi^\delta 
-\xi^\alpha\xi^\beta R_{\alpha\gamma\beta\delta}\Box\varphi^\gamma\Box\varphi^\delta
\nonumber\\
&&
+\xi^\alpha\xi^\beta
\left(\nabla_\alpha R_{\epsilon\gamma\beta\delta}+\nabla_\gamma R_{\epsilon\alpha\beta\delta}\right)
\partial_\mu\varphi^\gamma\partial^\mu\varphi^\delta \Box\varphi^\epsilon 
+\xi^\alpha\xi^\beta R_{\phi\gamma\delta\alpha}R^\phi\,\!_{\epsilon\eta\beta}
\partial_\mu\varphi^\gamma\partial^\mu\varphi^\delta
\partial_\nu\varphi^\epsilon\partial^\nu\varphi^\eta 
\nonumber\\
&&
+
2\nabla_\mu\xi^\alpha\nabla^\mu\xi^\beta \partial_\nu\varphi^\gamma\partial^\nu\varphi^\delta
T_{\alpha\beta\gamma\delta}
+4\nabla_\mu\xi^\alpha\nabla_\nu\xi^\beta \partial^\mu\varphi^\gamma\partial^\nu\varphi^\delta
T_{\alpha\gamma\beta\delta}
-2 \xi^\alpha\xi^\beta R^\phi{}_{\alpha\gamma\beta}T_{\phi\delta\epsilon\eta}
\partial_\mu\varphi^\gamma\partial^\mu\varphi^\delta
\partial_\nu\varphi^\epsilon\partial^\nu\varphi^\eta
\nonumber\\
&&
+4\xi^\alpha\nabla_\mu\xi^\beta\nabla_\alpha T_{\beta\gamma\delta\epsilon}
\partial^\mu\varphi^\gamma\partial_\nu\varphi^\delta\partial^\nu\varphi^\epsilon
+\frac{1}{2}\xi^\alpha\xi^\beta\nabla_\alpha\nabla_\beta T_{\gamma\delta\epsilon\eta}
\partial_\nu\varphi^\gamma\partial^\nu\varphi^\delta
\partial_\mu\varphi^\epsilon\partial^\mu\varphi^\eta
\Bigr]\ .
\end{eqnarray}
For notational simplicity here and in the following 
we drop the bars over $\varphi$, $\nabla$ and $R$, 
but it is always understood that they are computed at the background field.
The terms have been kept in the order in which they appear in \eqref{action}, namely the
first line comes from the variation of the two derivative term,
the second and third lines come from the variation of the term containing $\g4$
the last two lines come from the variation of the term containing $T$.

\subsection{The running effective action}

Our procedure for calculating the beta functions is a particular implementation of
Wilson's prescription that physics at the scale $k$ is described by an 
effective action $\Gamma_k$ where all modes with momenta $q>k$ have been integrated out.
We define formally an ``effective average action'' $\Gamma_k$ 
by implementing an {\it infrared} cutoff $k$ 
in the functional integral over the quantum field $\xi$.
If $\bar S(\varphi,\xi)$ is the bare action of the theory, 
the IR cutoff can be implemented by adding
to $\bar S$ a term $\Delta S_k(\varphi,\xi)$, quadratic in $\xi^\alpha$,
which in Fourier space would have the general structure:
\begin{equation}
\label{cutoff1}
\Delta S_k(\varphi,\xi)=\int d^4q\, \xi^\alpha(-q) R_{k\alpha\beta}(q^2)\xi^\beta(q)\ .
\end{equation}
The kernel $R_{k\alpha\beta}(q^2)$, sometimes also called the cutoff, 
is chosen in such a way that the propagation of 
field modes $\xi^\alpha(q)$ with $|q|<k$ is suppressed, while
field modes with $|q|>k$ are unaffected.
There is a vast freedom in the choice of the cutoff $\Delta S_k$, and in principle
physical predictions should turn out to be independent of this choice.
One can use this freedom to simplify calculations to some extent.
One possibility would be to write the cutoff exactly as in \eqref{cutoff1},
with $R_{k\alpha\beta}(q^2)=\g4_{\alpha\beta}R_k(q^2)$,
where $R_k(q^2)$ is a scalar function of the modulus of the momentum.
Note that $q^2$ is just the eigenvalue of the operator $-\partial^2$
acting on the quantum field.
It is more conveniente to write the cutoff in terms of the eigenvalues of some covariant
operator, such as the Laplacian constructed with the background field $-\nabla^2$.
This is the choice that was used in \cite{codello2}.
In this paper we will find it expedient to use instead of $-\nabla^2$
the full covariant fourth order operator 
$\Delta=\frac{\delta^2 S}{\delta\varphi\delta\varphi}$:
\begin{equation}
\label{cutoff2}
\Delta S_k(\varphi,\xi)=\frac{1}{2}\int d^4x\,
\xi^\alpha \g4_{\alpha\beta}(\varphi)R_k(\Delta)\xi^\beta\ .
\end{equation}
Because $\Delta$ depends only on the background field, and not on the quantum fields,
this cutoff is still quadratic in the quantum fields, as required.

Having modified the propagator of the theory, we define a generating functional $W_k(\varphi,j)$,
depending on the background $\varphi$ and on a source field $j$ coupled linearly to the quantum field $\xi$, by
\begin{equation}
\label{pathintegral}
W_k(\varphi,j)=-\log \int (d\xi^\alpha) \exp\left(-\bar S(\varphi,\xi)
-\Delta S_k(\varphi,\xi)-\int j_\alpha\xi^\alpha\right)
\end{equation}
Then we define a modified $k$--dependent Legendre transform
$$
\Gamma_k(\varphi,\xi)=W_k(\varphi,j)-\int j_\alpha\xi^\alpha -\Delta S_k(\varphi,\xi)\ ,
$$
where $\Delta S_k$ has been subtracted.
The ``classical fields'' $\frac{\delta W_k}{\delta j_\alpha}$ 
are denoted again $\xi^\alpha$ for notational simplicity.
The functional $\Gamma_k$ reduces for $k\to 0$ to the usual 
background field effective action $\Gamma(\varphi,\xi)$,
the generating functional of one--particle irreducible Green functions of $\xi$.

\subsection{The one loop beta functional}

At one loop one can evaluate the functional $\Gamma_k$:
\begin{equation}
\Gamma_k^{(1)}=S+\frac{1}{2}\mathrm{Tr}\log\left(\frac{\delta^2 S}{\delta\varphi\delta\varphi}+R_k\right)\ .
\end{equation}
Note that $\Delta S_k$ has cancelled out.
The only remaining dependence on $k$ is in $R_k$, so
\begin{equation}
\label{onelooperge}
k\frac{d\Gamma_k^{(1)}}{dk}=
\frac{1}{2}\mathrm{Tr}\left(\frac{\delta^2 S}{\delta\varphi\delta\varphi}+R_k\right)^{-1}
k\frac{dR_k}{dk}\ .
\end{equation}
The r.h.s. can be regarded as the one loop beta functional of the theory.
The individual beta functions can be read off by isolating the coefficients
of various operators.
Of course one could derive the one loop beta functions in other, more traditional ways.
We prefer this route because it has a few advantages.
First, due to the rapid fall off
of the function $k\partial_k R_k$, the beta functional is itself finite
and one does not actually need to introduce any ultraviolet regularization.
So, even though the derivation of the equation from a functional integral
was formal, because the functional integral is itself ill defined,
the functional RG equation is itself perfectly well defined.
A second important point is that the RG improvement of this equation,
where one replaces the bare action $S$ by $\Gamma_k$ in the r.h.s.,
is actually an exact equation \cite{wetterich}.
So although in the present work we shall restrict ourselves to the one loop approximation,
the formalism is ready for the calculation of the beta functions based on a truncation
of the exact RG equation, which amount to resumming infinitely many orders of perturbation theory.
A final, important point is that experience with other systems shows that this
procedure gives exactly the same results as any other procedure for the
universal (scheme-independent) one loop beta functions.
We will see in section IIID that, to the extent that a comparison is possible, this expectation
will be confirmed also in this case.

\subsection{Global symmetries}

If there are any symmetries, one can define the RG flow so as to preserve them.
To see this, let $\Phi$ be an internal symmetry, as in section IIA.
Since it is an isometry of $\g4$, it also leaves the connection invariant,
so it maps the geodesic through $y$ tangent to $\xi$
to the geodesic through $\Phi(y)$ tangent to $T\Phi(\xi)$ \cite{kn}:
$$
\Phi(Exp_y(\xi))=Exp_{\Phi(y)}(T\Phi(\xi))\ .
$$
We call $\varphi'=\Phi\circ\varphi$ 
and $\xi'=T\Phi(\xi)$ the transform of $\varphi$ and $\xi$ under $\Phi$.
Then
$\varphi'=\Phi(Exp_{\varphi}\xi)=Exp_{\Phi(\varphi)}(T\Phi(\xi))
=Exp_{\varphi'}\xi'$.
There follows that
\begin{equation}  
\bar S(\varphi',\xi')=S(\varphi')=S(\varphi)=\bar S(\varphi,\xi)\,,
\end{equation}
{\it i.e.} the background field action $\bar S$ is $G$-invariant provided both
background and quantum field are transformed.
The operator $\Delta$ is covariant, so
$ \Delta'(\xi')=T\Phi(\Delta(\xi))$
or abstractly $\Delta'=T\Phi\circ(\Delta)\circ T\Phi^{-1}$,
so also the cutoff term \eqref{cutoff2} is invariant:
\begin{equation}  
\Delta S_k(\varphi',\xi')=\Delta S_k(\varphi,\xi)\,,
\end{equation}
One can formally choose the measure in the functional
integral \eqref{pathintegral} to be invariant under $\Phi$.
Since both measure and integrand are invariant, the effective action $\Gamma_k$ will
also be invariant, for all $k$.

Somewhat less formally, one can arrive at the same conclusion as follows: 
observe that the cutoff as defined in \eqref{cutoff2} is a suppression term that depends
on the eigenvalue of the operator $\Delta$ on the normal modes of the field.
From the transformation properties of $\Delta$ one sees that
if $\xi$ is an eigenvector of $\Delta$ with eigenvalue $\lambda$,
then $\xi'$ is an eigenvector of $\Delta'$ with the same eigenvalue.
Therefore the spectrum of $\Delta$ is invariant.
Equation \eqref{onelooperge} gives the (one loop)
scale variation of $\Gamma_k(\varphi)$ as a sum of terms,
each term being a fixed function evaluated on an eigenvalues of $\Delta$.
Since the eigenvalues are invariant, the sum is also invariant,
so it follows that $\partial_t\Gamma_k(\varphi)$ is invariant.
This implies that if the starting action $\Gamma_{k_0}(\varphi)$ is invariant,
also the action at any other $k$ is.
This argument is mathematically more meaningful, because unlike the one
based on the path integral, it involves only statements about finite expressions.

The previous argument can be applied both to discrete and continuous symmetries.
For example in the case of discrete symmetries, it implies that the flow preserves Parity. 
If the $A$ term violates Parity, it must be set to zero in order to have a Parity invariant theory.
The flow will preserve this property, so the beta function of $A$ will be zero.
In other words the condition $A=0$ will be ``protected by Parity''.
We will see this in an explicit calculation in section IIIb.

\section{Evaluation of beta functions}

The one loop RG flow equation \eqref{onelooperge} can be approximated by resorting to a truncation,
which means keeping only a finite number of terms in $\Gamma_k$,
inserting this ansatz in the flow equation and deriving from it
the beta functions of the couplings that enter in the ansatz.
The best way of truncating $\Gamma_k$ is to do so consistently with a derivative expansion,
{\it i.e.} to keep all the terms with a given number of derivatives.
In this paper we will approximate $\Gamma_k$ by a functional of the form \eqref{action},
where the tensors $\g2$, $\g4$ and $T$ are $k$-dependent, and $A=0$.
In general this is still a functional flow, because the tensors
actually contain infinitely many couplings.
We will be able to say more in the case when a global symmetry restricts
the possible form of these tensors, so that only finitely many couplings remain.
In this paper we will explicitly compute the beta functions in the case when $Y$ is a 
sphere or a special unitary group. Since these are symmetric spaces,
it will be consistent to neglect the $A$ terms altogether.

\subsection{The inverse propagator}

Integrating by parts one can rewrite \eqref{secondvariation} in the form
$\bar S^{(2)}(\varphi,\xi)=\frac{1}{2}(\xi,\Delta\xi)$,
where the inner product of vectorfields along $\varphi$ is
$(\xi,\zeta)=\int d^4x\, \g4_{\alpha\beta}\xi^\alpha\zeta^\beta$
and $\Delta$ is a self-adjoint operator of the form:
\begin{equation}
\label{operator}
\Delta_{\alpha\beta} = 
\g4_{\alpha\beta}\Box^2 + {\cal B}^{\mu\nu}_{\alpha\beta}\nabla_\mu \nabla_\nu 
+ {\cal C}^\mu_{\alpha\beta}\nabla_\mu+{\cal D}_{\alpha\beta}\,. 
\end{equation}
Self-adjointness means that $(\xi,\Delta\zeta)=(\Delta\xi,\zeta)$ and implies the properties:
\begin{eqnarray}
{\cal B}^{\mu\nu}_{\alpha\beta}&=&{\cal B}^{\nu\mu}_{\beta\alpha}\ ,
\\
{\cal C}^\mu_{\alpha\beta}&=&-{\cal C}^\mu_{\beta\alpha}
+\nabla_\nu{\cal B}^{\mu\nu}_{\beta\alpha}
+\nabla_\nu{\cal B}^{\nu\mu}_{\beta\alpha}\ ,
\\
{\cal D}_{\alpha\beta}&=&{\cal D}_{\beta\alpha}
+\nabla_\nu{\cal C}^\mu_{\beta\alpha}
+\nabla_\nu\nabla_\nu{\cal B}^{\nu\mu}_{\beta\alpha}\ .
\end{eqnarray}
In addition by commuting derivatives we can arrange the operator so that 
${\cal B}^{\mu\nu}_{\alpha\beta}={\cal B}^{\nu\mu}_{\alpha\beta}$.
In order to arrive at the operator $\Delta$ we proceed in two steps.
First we put all the derivatives of \eqref{secondvariation} on one of the $\xi$'s, so that 
$\bar S^{(2)}(\varphi,\xi)=\frac{1}{2}(\xi,\tilde\Delta\xi)$,
where $\tilde\Delta$ is of the form \eqref{operator},  with
\begin{eqnarray}
\label{bi}
\tilde{\cal B}^{\mu\nu}_{\alpha\beta}&=&\delta^{\mu\nu}
(-h^{(2)}_{\alpha\beta}
+2\partial_\rho\varphi^\gamma\partial^\rho\varphi^\delta
(R_{\alpha\gamma\beta\delta}
- T_{\alpha\beta\gamma\delta}))
-4\partial_\mu\varphi^\gamma\partial_\nu\varphi^\delta
T_{\alpha\gamma\beta\delta}\ ,
\\
\tilde{\cal D}_{\alpha\beta}&=&
\partial_\mu\varphi^\gamma\partial^\mu\varphi^\delta
\left(\frac{1}{2}\nabla_\alpha\nabla_\beta\g2_{\gamma\delta}
-\g2_{\gamma\epsilon}R^\epsilon{}_{\beta\delta\alpha}\right)
-\Box\varphi^\gamma\Box\varphi^\delta R_{\alpha\gamma\beta\delta}
-2 \partial_\rho\varphi^\gamma\partial^\rho\varphi^\delta\Box\varphi^\epsilon
\nabla_{(\delta}R_{\alpha)\epsilon\beta\gamma}
\nonumber\\
&&
+\partial_\rho\varphi^\gamma\partial^\rho\varphi^\delta
\partial_\sigma\varphi^\epsilon\partial^\sigma\varphi^\eta
\left(
R_{\alpha\gamma\delta\phi}R_{\beta\epsilon\eta}{}^\phi
+\frac{1}{2}\nabla_\alpha\nabla_\beta T_{\gamma\delta\epsilon\eta}
+2R_{\phi\alpha\beta\epsilon}T^\phi{}_{\eta\gamma\delta}\right)
\ .
\end{eqnarray}
We do not display the form of $\tilde{\cal C}^\mu_{\alpha\beta}$,
since it does not contribute to the expressions we want to calculate,
as will become clear in due course.
This operator $\tilde\Delta$ is not self-adjoint, and we define 
$\Delta={\tiny\frac{1}{2}}(\tilde\Delta+\tilde\Delta^\dagger)$.
Its coefficients are
\begin{eqnarray}
{\cal B}^{\mu\nu}_{\alpha\beta}&=&
\frac{1}{2}\left(\tilde{\cal B}^{\mu\nu}_{\alpha\beta}+\tilde{\cal B}^{\nu\mu}_{\beta\alpha}\right)\ ,
\nonumber\\
{\cal C}^\mu_{\alpha\beta}&=&\frac{1}{2}\left(\tilde{\cal C}^\mu_{\alpha\beta}-\tilde{\cal C}^\mu_{\beta\alpha}
+\nabla_\nu\tilde{\cal B}^{\mu\nu}_{\beta\alpha}
+\nabla_\nu\tilde{\cal B}^{\nu\mu}_{\beta\alpha}\right)\ ,
\nonumber\\
{\cal D}_{\alpha\beta}&=&\frac{1}{2}\left(\tilde{\cal D}_{\alpha\beta}
+\tilde{\cal D}_{\beta\alpha}
-\nabla_\mu\tilde{\cal C}^\mu_{\beta\alpha}
+\nabla_\mu\nabla_\nu\tilde{\cal B}^{\nu\mu}_{\beta\alpha}
\right)\ .
\end{eqnarray}
Note that the last two terms in ${\cal C}^\mu_{\alpha\beta}$
and ${\cal D}^\mu_{\alpha\beta}$ are total derivatives,
and will not contribute to our final formulae.
Finally we can symmetrize ${\cal B}^{\mu\nu}_{\alpha\beta}$
in $\mu$, $\nu$ at the cost of generating a commutator term that
contributes to ${\cal D}_{\alpha\beta}$.
The final form of the operator $\Delta$ is \eqref{operator}, with 
\begin{eqnarray}
{\cal B}^{\mu\nu}_{\alpha\beta}&=&\delta^{\mu\nu}
(-h^{(2)}_{\alpha\beta}
+2\partial_\rho\varphi^\gamma\partial^\rho\varphi^\delta
(R_{\alpha\gamma\beta\delta}
- T_{\alpha\beta\gamma\delta}))
-2\partial_\mu\varphi^\gamma\partial_\nu\varphi^\delta
(T_{\alpha\gamma\beta\delta}+T_{\alpha\delta\beta\gamma})\ ,
\nonumber\\
{\cal D}_{\alpha\beta}&=&\frac{1}{2}\left(\tilde{\cal D}_{\alpha\beta}
+\tilde{\cal D}_{\beta\alpha}\right)
-\partial_\rho\varphi^\gamma\partial^\rho\varphi^\delta
\partial_\sigma\varphi^\epsilon\partial^\sigma\varphi^\eta
(T_{\alpha\gamma\epsilon\phi}R_{\delta\eta\beta}{}^\phi
+T_{\beta\gamma\epsilon\phi}R_{\delta\eta\alpha}{}^\phi)
+\mathrm{TD}\ ,
\end{eqnarray}
where TD stands for ``total derivatives''.
Again we omit to give ${\cal C}^\mu_{\alpha\beta}$, because it 
does not contribute to the beta functions.
These formulae agree with (3.17-21) in \cite{bk}, except for a factor 2 in the coefficient
of the first term containing $T_{\alpha\beta\gamma\delta}$ in equation \eqref{bi}.

\subsection{Beta functionals}

We begin by discussing the general case of the action \eqref{action}
with arbitrary $h^{(2)}$, $h^{(4)}$ and $T$, and $A=0$.
We evaluate the trace in \eqref{onelooperge} by heat kernel methods.
The advantage of this procedure is that pieces of the
calculation are readily available in the literature.
Given a differential operator $\Delta$ of order $p$, 
and some function $W$, we have
\begin{equation}
\label{HKE}
Tr W(\Delta) = \frac{1}{(4\pi)^{2}}\Bigl[Q_{\frac{4}{p}}(W)B_0(\Delta)+
Q_{\frac{2}{p}}(W)B_2(\Delta)+ Q_{0}(W)B_4(\Delta)+\ldots\Bigr]\,.
\end{equation}
The heat kernel coefficients are defined by the asymptotic expansion
\begin{equation}
\label{heatkernel}
{\rm Tr} (e^{-s\Delta})=
\frac{1}{(4\pi)^2}
\left[B_0s^{-4/p}+B_2s^{-2/p}+B_0+\ldots\right]\ ,
\end{equation}
with $B_n=\int d^4x \mathrm{tr} b_n$; $b_n$ are matrices with indices
$\alpha$, $\beta$ and tr denotes the trace over such indices.
For a fourth order operator of the form
\eqref{operator}, they can be found in \cite{barth}.
The quantities $Q_n(W)$ in \eqref{HKE} are given by
$Q_n(W) = \frac{1}{\Gamma(n)} \int_0^\infty dz z^{n-1} W(z)$
for $n>0$ and $Q_0(W)=W(0)$.
We do not need any higher coefficients.
In order to be able to evaluate the integrals in closed form we choose the
``optimized'' cutoff function $R_k(z)=(k^4-z)\theta(k^4-z)$ \cite{optimized}.
The scale derivative of the cutoff is
${\tiny k\frac{dR_k}{dk}}=4k^4\theta(k^4-z)$, and the
modified inverse propagator $P_k(z)=z+R_k(z)$ is equal to $k^4$ for $z<k^4$.
Then the function to be traced in the ERGE is just a step function:
$W(z)=
\frac{1}{2}\frac{1}{P_k}k\frac{dR_k}{dk}=2\theta(1-z/k^4)$,
and the integrals are very simple:
\begin{equation}
Q_1 = 2 k^4\,,\qquad
Q_{\frac{1}{2}}=\frac{4}{\sqrt{\pi}}k^2\,,\qquad
Q_0 = 2\,.
\end{equation}
The first term in \eqref{HKE} is field independent and will be omitted.
Putting together the remaining pieces:
\begin{equation}
k\frac{d\Gamma_k}{dk}=
\frac{1}{(4\pi)^{2}}\int d^4x
\left(
\frac{1}{4}k^2{\cal B}^{\alpha}_{\alpha}
+\frac{1}{6}\Omega_{\mu\nu}^{\alpha\beta}\Omega^{\mu\nu}_{\beta\alpha}
+\frac{1}{24}{\cal B}^{\alpha\beta}_{\mu\nu}{\cal B}^{\mu\nu}_{\beta\alpha}
+\frac{1}{48}{\cal B}_{\alpha\beta}{\cal B}^{\beta\alpha}
-{\cal D}^\alpha_\alpha
\right)
\end{equation}
where $\Omega$ is defined as in \eqref{curvature} and ${\cal B}={\cal B}^{\mu}_{\mu}$.
The first term comes from $B_2$, the others from $B_4$.
One finds
\begin{eqnarray}
\label{traceb}
\frac{1}{4}{\cal B}^{\alpha}_{\alpha}&=&
\partial_\mu\varphi^\gamma\partial^\mu\varphi^\delta
\left(2R_{\gamma\delta}-2T^\alpha{}_{\alpha\gamma\delta}
-T^\alpha{}_{\gamma\alpha\delta}\right)
\\
\label{omegasquare}
\frac{1}{6}\Omega_{\mu\nu}^{\alpha\beta}\Omega^{\mu\nu}_{\beta\alpha}&=&
-\frac{1}{6}\partial_\mu\varphi^\alpha\partial^\mu\varphi^\beta
\partial_\nu\varphi^\gamma\partial^\nu\varphi^\delta
R_{\alpha\gamma\epsilon\eta}R_{\beta\delta}{}^{\epsilon\eta}
\\
\label{tracebb}
\frac{1}{24}{\cal B}^{\alpha\beta}_{\mu\nu}{\cal B}^{\mu\nu}_{\beta\alpha}
+\frac{1}{48}{\cal B}_{\alpha\beta}{\cal B}^{\beta\alpha}&=&
\frac{1}{2}\g2_{\alpha\beta}{\g2}^{\alpha\beta}
+\partial_\mu\varphi^\alpha\partial^\mu\varphi^\beta
\left(T_\alpha{}^\gamma{}_\beta{}^\delta
+2T_{\alpha\beta}{}^{\gamma\delta}
-2 R_\alpha{}^\gamma{}_\beta{}^\delta\right)
\g2_{\gamma\delta}
\nonumber\\
&&
+\partial_\mu\varphi^\alpha\partial^\mu\varphi^\beta
\partial_\nu\varphi^\gamma\partial^\nu\varphi^\delta
\Bigl[
\frac{2}{3}T_{\alpha\epsilon\gamma\eta}T_\beta{}^{(\epsilon}{}_\delta{}^{\eta)}
+\frac{1}{3}T_{\alpha\epsilon\beta\eta}T_\gamma{}^{(\eta}{}_\delta{}^{\epsilon)}
+4T_{\alpha(\epsilon\gamma)\eta}T_\beta{}^{(\epsilon}{}_\delta{}^{\eta)}
\nonumber\\
&&
-4R_{\alpha\epsilon\gamma\eta}T_\beta{}^{(\epsilon}{}_\delta{}^{\eta)}
-2R_{\alpha\epsilon\beta\eta}T_\gamma{}^{(\eta}{}_\delta{}^{\epsilon)}
+2R_{\alpha\epsilon\beta\eta}R_\gamma{}^{(\eta}{}_\delta{}^{\epsilon)}
\Bigr]
\\
\label{traced}
-{\cal D}^\alpha_\alpha&=&
\Box\varphi^\alpha\Box\varphi^\beta R_{\alpha\beta}
+\Box\varphi^\alpha\partial^\mu\varphi^\beta
\partial_\mu\varphi^\gamma
(2\nabla_\gamma R_{\alpha\beta}-\nabla_\alpha R_{\beta\gamma})
\nonumber\\
&&
+\partial_\mu\varphi^\alpha\partial^\mu\varphi^\beta
\left(\g2_{\alpha\gamma}R^\gamma{}_\beta
-\frac{1}{2}\nabla_\gamma\nabla^\gamma\g2_{\alpha\beta}\right)
+\partial_\mu\varphi^\alpha\partial^\mu\varphi^\beta
\partial_\nu\varphi^\gamma\partial^\nu\varphi^\delta\times
\nonumber\\
&&
\Bigl(
2R_\alpha{}^\epsilon T_{\epsilon\beta\gamma\delta}
+2R_{\beta\delta}{}^{\epsilon\eta}T_{\alpha\eta\gamma\epsilon}
-\frac{1}{2}\nabla_\epsilon\nabla^\epsilon T_{\alpha\beta\gamma\delta}
-R_{\alpha\epsilon\beta\eta}R_\gamma{}^\epsilon{}_\delta{}^\eta
\Bigr)
\end{eqnarray}
From here one can read off the beta functionals of $\g2$, $A$, $T$
as the coefficients of terms containing two, three and four
powers of $\partial_\mu\varphi^\alpha$, respectively.
We do not give these general formulae, but just make some observations.
The only term proportional to $\Box\varphi^\alpha\Box\varphi^\beta$ is contained
in $-{\cal D}^\alpha_\alpha$, so the beta functional of $h^{(4)}$
is easily obtained:
\begin{equation}
k\frac{d}{dk}h^{(4)}_{\alpha\beta}=\frac{1}{8\pi^2}R_{\alpha\beta}\,.
\end{equation}
This is very similar to the result for the two derivative truncation.
In order to compare results obtained with the same type of cutoff, we should
repeat the calculation of \cite{codello2} using a cutoff constructed with the
full inverse propagator 
$\Delta_{\alpha\beta}=-\g2_{\alpha\beta}\nabla^2-
\partial_\mu\varphi^\gamma\partial_\mu\varphi^\delta R_{\alpha\gamma\beta\delta}$.
This is a cutoff of type III in the terminology used in \cite{cpr}.
In this case the general beta function of the metric is
\begin{equation}  
\label{betatrunc2III}
k\frac{d}{dk}\g2_{\alpha\beta}=
\frac{1}{(4\pi)^2} Q_1\left(\frac{\dot R_k}{P_k}\right)R_{\alpha\beta}
=\frac{1}{8\pi^2}k^2 R_{\alpha\beta}\ ,
\end{equation}
where $R$ denotes now the curvature of $\g2_{\alpha\beta}$.
As a side remark, this little calculation is also useful to test the scheme dependence 
of the results: with the type I cutoff used in \cite{codello2} the result was
\begin{equation}  
\label{betatrunc2I}
k\frac{d}{dk}\g2_{\alpha\beta}
=\frac{1}{(4\pi)^2} Q_2\left(\frac{\dot R_k}{P_k^2}\right)R_{\alpha\beta}
=\frac{1}{16\pi^2} k^2 R_{\alpha\beta}\ ,
\end{equation}
which differs by a factor 2.

Another fact that follows from \eqref{traced} is that the beta function
of $A$ (coming from the coefficient of 
$\Box\varphi^\alpha\partial^\mu\varphi^\beta\partial_\mu\varphi^\gamma$)
is proportional to covariant derivatives of the Ricci tensor.
For symmetric spaces the covariant derivative of the curvature vanishes
and therefore on such spaces it is consistent to set $A=0$.
This confirms the general statement made in section IIE.
The particular models that we shall consider in the following are
symmetric spaces.

\subsection{The spherical models}

We now consider the class of models for which the target space $Y$ is the sphere $S^n$.
Such models are often called the $O(N)$ models, with $N=n+1$, because they have global 
symmetry $O(N)$.
There is only one $O(n+1)$-invariant nonvanishing rank two tensor on the sphere, 
there is no invariant rank three tensor and
there are only two invariant rank four tensors with the desired index symmetries,
up to overall constant factors.
If we regard $S^n$ as embedded in $\mathbf{R}^{n+1}$, we call $h_{\alpha\beta}$ the metric
of the sphere of unit radius.
Its Riemann and Ricci tensors are given by
$$
R_{\alpha\beta\gamma\delta}=h_{\alpha\gamma}h_{\beta\delta}-h_{\alpha\delta}h_{\beta\gamma}\ ;\quad
R_{\alpha\beta}=(n-1)h_{\alpha\beta}\ ;\quad
R=n(n-1)\ .
$$
Therefore both $\g2$ and $\g4$ must be proportional to $h$, and $T$ is a combination of $h$'s:
$$
\g2_{\alpha\beta}=\frac{1}{g^2}h_{\alpha\beta}\ ;\qquad
\g4_{\alpha\beta}=\frac{1}{\lambda}h_{\alpha\beta}\ ;\qquad
T_{\alpha\beta\gamma\delta}=\frac{\ell_1}{2}\left(h_{\alpha\gamma}h_{\beta\delta}+h_{\alpha\delta}h_{\beta\gamma}\right)
+\ell_2 h_{\alpha\beta}h_{\gamma\delta}\ .
$$
Here $g^2$ has mass dimension $2$, while $\lambda$, $\ell_1$, $\ell_2$ are dimensionless
\footnote{the names $\ell_1$ and $\ell_2$ are used commonly in chiral perturbation theory \cite{gl}.}.
It is convenient to regard $\tiny\frac{1}{\lambda}$ as the overall factor of the fourth order terms;
then we define the ratios between the three coefficients of the four-derivative terms as 
$f_1=\lambda\ell_1$ and $f_2=\lambda\ell_2$.
For the reader's convenience we rewrite the action of the $S^n$ models:
\begin{equation}
\label{sphericalaction}
\int d^4x\Biggl[
\frac{1}{2g^2}h_{\alpha\beta}\partial_\mu\varphi^\alpha\partial^\mu\varphi^\beta
+\frac{1}{2\lambda}\left(h_{\alpha\beta}\Box\varphi^\alpha\Box\varphi^\beta
+\partial_\mu\varphi^\alpha\partial^\mu\varphi^\beta\partial_\nu\varphi^\gamma\partial^\nu\varphi^\delta
(f_1h_{\alpha\gamma}h_{\beta\delta}+f_2h_{\alpha\beta}h_{\gamma\delta})\right)
\Biggr]
\end{equation}
One then finds the following beta functions:
\begin{eqnarray}
\label{bls}
\beta_\lambda & = & -\frac{n-1}{8\pi^2}\lambda^2
\\
\label{b1s}
\beta_{f_1} & = &\frac{\lambda}{48\pi^2}
\left((n+21)f_1^2+20f_2f_1
+4f_2^2+6(n+3)f_1+24f_2+8\right)
\\
\label{b2s}
\beta_{f_2} & = &\frac{\lambda}{8\pi^2}
\left(\frac{n+15}{12}f_1^2+\frac{3n+17}{3}f_1f_2
+\frac{6n+7}{3}f_2^2-(n+3)f_1-(3n+1)f_2+n-\frac{7}{3}\right)
\\
\label{bgs}
\beta_{\tilde g^2} & = & 2\tilde g^2
+\frac{\tilde g^4}{16\pi^2}\left((5+n)f_1+(2+4n)f_2+4(1-n)\right)
-\frac{\lambda\tilde g^2}{16\pi^2}\left((5+n)f_1+(2+4n)f_2+2(1-n)\right)\phantom{aaa}
\end{eqnarray}
Equations \eqref{b1s} and \eqref{b2s} differ in a significant way from equation
(5.11) in \cite{bk}. This is due to the already mentioned factor 2 in a term in \eqref{bi}. 
Unfortunately, this changes completely the picture of the fixed points.

It will be instructive to compare the results of this four derivative truncation
with those of the simpler two derivative truncation discussed in \cite{codello2}.
If we specialize \eqref{betatrunc2III} to $Y=S^n$, it gives
\begin{equation}  
\label{betatrunc2sphere}
k\frac{d\tilde g^2}{dk}=2\tilde g^2-\frac{n-1}{8\pi^2}\tilde g^4\,,
\end{equation}
whereas from \eqref{bgs}, setting for simplicity
$\ell_1=\ell_2=0$ and in the limit $\lambda\to 0$ one gets
\begin{equation}  
\label{betatrunc4sphere}
k\frac{d\tilde g^2}{dk}=2\tilde g^2-\frac{n-1}{4\pi^2}\tilde g^4\,.
\end{equation}
The difference is just a factor $2$, which is within the
range of variation due to the scheme dependence.
It is quite remarkable that the beta function 
is so similar in spite of the very different dynamics.
We shall see in section IVA that this fact is quite general.

\subsection{The chiral models}

Next we consider the case where $Y$ is the group $SU(N)$.
In this case it is customary to denote $U(x)$ the matrix (in the fundamental representation) 
that corresponds to the coordinates $\varphi^\alpha$.
We demand that the theory be invariant under left and right multiplications
$U(x)\mapsto g_L^{-1}U(x)g_R$, forming the group $SU(N)_L\times SU(N)_R$
(``chiral symmetry'').
Further we demand that the theory be invariant under the discrete symmetries
$U(x)\mapsto U^T(x)$, which corresponds physically to charge conjugation,
to the simple parity $x_1\mapsto-x_1$, to the involutive isometry
$\Phi_0:U\to U^{-1}$ 
and hence to Parity $U(x_1,x_2,x_3,x_4)\mapsto U^{-1}(-x_1,x_2,x_3,x_4)$.
More details on the translation between the tensor and the matrix formalism
are given in Appendix A.

Let $e_a$ be a basis of the Lie algebra, with $a=1\ldots n^2-1$.
We denote $T_a$ the corresponding matrices in the fundamental representation; 
they are a set of hermitian, traceless $N\times N$ matrices.
We fix the normalization of the basis by the equation
\begin{equation}
\label{normalization}
T_aT_b=\frac{1}{2N}\delta_{ab}+\frac{1}{2}\left(d_{abc}+if_{abc}\right)T_c\,.
\end{equation}
(In the case of $SU(3)$ these matrices are one half the Gell-Mann $\lambda$ matrices.)

A tensor on $SU(N)$ which is invariant under 
$SU(N)_L\times SU(N)_R$ is said to be ``biinvariant''.
There is a one to one correspondence between
biinvariant tensors on $SU(N)$ and $Ad$-invariant tensors
in the Lie algebra of $SU(N)$, where $Ad$ is the adjoint representation.
Given an $Ad$-invariant tensor $t_{ab\ldots}{}^{cd\ldots}$ on the algebra, the
corresponding biinvariant tensorfield on the group is
$$
t_{\alpha\beta\ldots}{}^{\gamma\delta\ldots}=
t_{ab\ldots}{}^{cd\ldots}L^a_\alpha L^b_\beta\ldots L_c^\gamma L_d^\delta\ldots
$$
where $L^a_\alpha$ are the components of the left-invariant Maurer Cartan form
$L=U^{-1}dU=L^a_\alpha dy^\alpha (-iT_a)$ and $L_a^\alpha$ are the components of the left-invariant
vectorfields on $SU(N)$. The matrix $L_a^\alpha$ is the inverse of $L^a_\alpha$.
(In this construction we could use equivalently right-invariant objects.)

Up to rescalings, there is a unique $Ad$-invariant inner product in the Lie algebra,
which we choose as $h_{ab}=2\mathrm{Tr}T_aT_b=\delta_{ab}$
\footnote{Here the matrices are in the fundamental representation.
The Cartan-Killing form just differs by a constant: $B_{ab}=\mathrm{Tr}(Ad(T_a)Ad(T_b))=N\delta_{ab}$.}.
Then the corresponding biinvariant metric is
\begin{equation}
\label{biinvariantmetric}
h_{\alpha\beta}=L^a_\alpha L^b_\beta\delta_{ab}\ ,
\end{equation}
so that the left-invariant vectorfields $L_a$ can also be regarded as a vierbein.
The Riemann and Ricci tensors and the Ricci scalar of $h$ are given by
\begin{equation}
\label{curvaturesun}
R_{\alpha\beta\gamma\delta}=\frac{1}{4}L^a_\alpha L^b_\beta L^c_\gamma L^d_\delta f_{ab}{}^e f_{ecd}\ ;\qquad
R_{\alpha\beta}=\frac{1}{4}N h_{\alpha\beta}\ ;\qquad
R=\frac{1}{4}N(N^2-1)\ .
\end{equation}
As with the sphere, we define
$\g2_{\alpha\beta}=\frac{1}{g^2}h_{\alpha\beta}$, 
$\g4_{\alpha\beta}=\frac{1}{\lambda}h_{\alpha\beta}$.
The tensors $d_{abc}$ and $f_{abc}$ are a totally symmetric and a
totally antisymmetric $Ad$-invariant three tensor in the algebra.
In principle chiral invariance would permit a term in the action with
$A_{\alpha\beta\gamma}=L^a_\alpha L^b_\beta L^c_\gamma d_{abc}$;
however using $L^a_\alpha(\Phi_0(y))=R^a_\alpha(y)$,
$L_a^\alpha(y)R^b_\alpha(y)=Ad(g(y))^b{}_a$
and the $Ad$-invariance of $d_{abc}$, one sees that
$A_{\alpha\beta\gamma}(\Phi_0(y))=A_{\alpha\beta\gamma}(y)$,
so this term violates Parity.

For $T$ we have the following $Ad$-invariant
four-tensors in the algebra with the correct symmetries:
\begin{eqnarray}
T^{(1)}_{abcd}&=&\frac{1}{2}\left(\delta_{ac}\delta_{bd}+\delta_{ad}\delta_{bc}\right)\ ;\qquad
T^{(2)}_{abcd}=\delta_{ab}\delta_{cd}\ ;\qquad
T^{(3)}_{abcd}=\frac{1}{2}\left(f_{ace}f_{bd}{}^e+f_{ade}f_{bc}{}^e\right)\ ;
\nonumber
\\
T^{(4)}_{abcd}&=&\frac{1}{2}\left(d_{ace}d_{bd}{}^e+d_{ade}d_{bc}{}^e\right)\ ;\qquad
T^{(5)}_{abcd}=d_{abe}d_{cd}{}^e\ .
\end{eqnarray}
They are not all independent, however.
The identity (2.10) of \cite{macfarlane} implies that
\begin{equation}
\label{firstrelation}
\frac{2}{N}T^{(1)}-\frac{2}{N}T^{(2)}+T^{(3)}+T^{(4)}-T^{(5)}=0\ ,
\end{equation}
so that $T^{(5)}$ can be eliminated.
In the case $N=3$ the identity (2.23) of \cite{macfarlane},
together with the preceding relation, further implies
\begin{equation}
\label{secondrelation}
T^{(2)}-T^{(3)}-3 T^{(4)}=0\ ,
\end{equation}
so that we can also eliminate $T^{(4)}$.
Finally in the case $N=2$ the tensor $d_{abc}$ is identically zero,
so we can keep only $T^{(1)}$ and $T^{(2)}$ as independent combinations,
and use $T^{(3)}=T^{(2)}-T^{(1)}$.

The action of the generic $SU(N)$ models can then be written in the form:
\begin{equation}
\label{actionsun}
\int d^4x\Biggl[
\frac{1}{2g^2}h_{\alpha\beta}\partial_\mu\varphi^\alpha\partial^\mu\varphi^\beta
+\frac{1}{2\lambda}h_{\alpha\beta}\Box\varphi^\alpha\Box\varphi^\beta
+\frac{1}{2}\partial_\mu\varphi^\alpha\partial^\mu\varphi^\beta
\partial_\nu\varphi^\gamma\partial^\nu\varphi^\delta
\sum_{i=1}^4 \ell_i T^{(i)}_{\alpha\beta\gamma\delta}
\Biggr]
\end{equation}
and the sum stops at $i=3$ and $i=2$ for $N=3$ and $N=2$ respectively.
As in \eqref{sphericalaction}, it will be convenient to use instead of the couplings $\ell_i$
the combinations $f_i=\lambda\ell_i$.

Making repeated use of traces given in \cite{macfarlane2} one finds the following beta functions:
\begin{eqnarray}
\label{bsunlambda}
\beta_\lambda & = & -\frac{N}{32\pi^2}\lambda^2
\\
\label{bsun1}
\beta_{f_1} & = &
\frac{\lambda}{768\pi^2 N^2}
\Bigl[16N^2(N^2+20)f_1^2+64N^2f_2^2+180N^2f_3^2+4(149N^2-1280)f_4^2
\nonumber\\
&&
\qquad\qquad
+320N^2f_1f_2-32N^3f_1f_3+32N(N^2+4)f_1f_4+128Nf_2f_4-120N^2f_3f_4
\nonumber\\
&&
\qquad\qquad
+24N^3f_1-108N^2f_3+36N^2f_4+9N^2
\Bigr]
\\
\label{bsun2}
\beta_{f_2} & = &
\frac{\lambda}{768\pi^2 N^2}
\Bigl[
8N^2(N^2+14)f_1^2+32N^2(6N^2+1)f_2^2+60N^2f_3^2+4(7N^2+656)f_4^2
\nonumber\\
&&
\qquad\qquad
+32N^2(3N^2+14)f_1f_2+80N^3f_1f_3+16N(7N^2-44)f_1f_4+288N^3f_2f_3
\\
&&
\qquad\qquad
+32N(15N^2-64)f_2f_4+120N^2f_3f_4-24N^3(f_1+3f_2)-36N^2(f_3+f_4)+3N^2
\Bigr]
\nonumber\\
\label{bsun3}
\beta_{f_3} & = &
\frac{\lambda}{1536\pi^2 N}
\Bigl[52N^2f_3^2+12(23N^2-320)f_4^2+768Nf_1f_3+256Nf_1f_4
+384Nf_2f_3+128Nf_2f_4
\nonumber\\
&&
\qquad\qquad
+24(11N^2-64)f_3f_4-192N(f_1+f_2)-60N^2(f_3+f_4)+384f_4+N^2
\Bigr]
\\
\label{bsun4}
\beta_{f_4} & = &
\frac{\lambda}{1536\pi^2 N}
\Bigl[60N^2f_3^2+4(87N^2-1728)f_4^2+1536Nf_1f_4
+768Nf_2f_4+216N^2f_3f_4
\nonumber\\
&&
\qquad\qquad
-36N^2(f_3+f_4)+3N^2
\Bigr]
\\
\label{bsung}
\beta_{\tilde g^2} & = & 2\tilde g^2
+\frac{\tilde g^4}{16 N\pi^2}
\left(N(N^2+4)f_1+2N(2N^2-1)f_2+3N^2f_3+5(N^2-4)f_4-N^2\right)
\nonumber\\
&&
-\frac{\lambda\tilde g^2}{16 N\pi^2}
\left(N(N^2+4)f_1+2N(2N^2-1)f_2+3N^2f_3+5(N^2-4)f_4-N^2/2\right)
\qquad
\end{eqnarray}
In appendix A we establish the dictionary between our notation and that
used in \cite{hasenfratz}. When the beta functions are compared, we find
perfect agreement, except for one small difference:
the very last term in the first line of $\beta_{\tilde g^2}$ would be $N^2/2$ according to
\cite{hasenfratz}, {\it i.e.} $\tilde g^4$ and $\lambda\tilde g^2$
would have the same coefficients.
This is the same difference that we observed between \eqref{betatrunc2III}
(type III cutoff) and \eqref{betatrunc2I} (type I cutoff),
so, effectively the calculation in \cite{hasenfratz} is equivalent to a type I cutoff.
Given that the calculation in \cite{hasenfratz} was done using completely
different techniques, this agreement confirms that the one loop beta functions
of the dimensionless couplings (which in a calculation of the effective action
would correspond to logarithmic divergences) is scheme independent.

The cases $N=3$ and $N=2$ have to be treated separately, because in these
cases only three, respectively two, of the couplings $f_i$ are independent.
In the case $N=3$ one can eliminate $f_4$ in favor of the other three couplings.
Then using \eqref{secondrelation}
one can obtain the beta functions of $f_1$, $f_2$ and $f_3$
from the ones given above by
\begin{eqnarray*}
\label{betasuthree}
\beta_{f_1}\Big|_{N=3} & = &\beta_{f_1}\Big|_{N=3,f_4=0}
=\frac{\lambda}{768\pi^2}
\left[464f_1^2+64f_2^2+180f_3^2+320f_1f_2-96f_1f_3+72f_1-108f_3+9\right]
\\
\beta_{f_2}\Big|_{N=3} & = &\beta_{f_2}+\frac{1}{3}\beta_{f_4}\Big|_{N=3,f_4=0}
\\
= \frac{\lambda}{1536\pi^2}&&\!\!\!\!\!\!\!\!
\Bigl[368f_1^2+3520f_2^2+180f_3^2+2624f_1f_2+480f_1f_3+1728f_2f_3-144f_1-432f_2-108f_3+9\Bigr]
\\
\beta_{f_3}\Big|_{N=3} &=&\beta_{f_3}-\frac{1}{3}\beta_{f_4}\Big|_{N=3,f_4=0}
=\frac{\lambda}{32\pi^2}
\left[2f_3^2+16f_1f_3+8f_2f_3-4f_1-4f_2-3f_3\right]
\end{eqnarray*}

In the case $N=2$ we can set $f_4=0$, because $T^{(4)}=0$ identically,
and we can eliminate $f_3$.
One can obtain the beta functions of $f_1$, $f_2$ from the ones given above by
\begin{eqnarray*}
\label{betasutwo}
\beta_{f_1}\Big|_{N=2} & = &\beta_{f_1}-\beta_{f_3}\Big|_{N=2,f_3=0,f_4=0}
=\frac{\lambda}{96\pi^2}
\left[48f_1^2+8f_2^2+40f_1f_2+18f_1+12f_2+1\right]
\\
\beta_{f_2}\Big|_{N=2} & = &\beta_{f_2}+\beta_{f_3}\Big|_{N=2,f_3=0,f_4=0}
=\frac{\lambda}{192\pi^2}
\left[36f_1^2+200f_2^2+208f_1f_2-36f_1-60f_2+1\right]
\end{eqnarray*}
The latter result can be used to check also our beta functions for the spherical sigma model.
In fact there is exactly one manifold which is simultaneously a sphere
and a special unitary group: it is $SU(2)=S^3$.
Thus the beta functions should agree in this case. 
Before comparing, a little point needs
to be addressed. In section IIIC we chose the metric $h_{\alpha\beta}$
to be that of a sphere of unit radius.
In this section we have fixed the metric by the conditions \eqref{normalization},
\eqref{biinvariantmetric}.
It turns out that in the case $N=2$ this normalization corresponds to a sphere of radius two.
This can be seen for example from equation \eqref{curvaturesun}, specialized
to $N=2$, with $f_{abc}=\varepsilon_{abc}$.
In order to compare the beta functions of $S^3$
with those for $SU(2)$ we therefore have to redefine $\lambda\to\lambda/4$,
$f_1\to 4f_1$, $f_2\to 4f_2$, $g^2\to g^2/4$.
With these redefinitions, the beta functions do indeed agree.

\section{Fixed points}

\subsection{The spherical models}

We now discuss solutions of the RG flow equations.
The beta function of $\lambda$ depends only on $\lambda$ and the solution is
\begin{equation}
\label{lambdasol}
\lambda(t)=\frac{\lambda_0}{1+\lambda_0\frac{n-1}{8\pi^2}(t-t_0)}\ ,
\end{equation}
where $\lambda_0=\lambda(t_0)$. We assume $\lambda_0>0$, thus $\lambda$ is asymptotically free.
The beta functions of $f_1$ and $f_2$ do not depend on $g$,
so their flow can be studied independently.
Here we do not discuss general solutions but merely look for fixed points.
The overall factor $\lambda$ in these beta functions can be eliminated by
a simple redefinition $t=t(\tilde t)$ of the parameter along the RG trajectories:
\begin{equation}
\label{redef}
\frac{d}{d\tilde t}=\frac{1}{\lambda}\frac{d}{dt}\,.
\end{equation}
Since $\tilde t$ is a monotonic function of $t$,
the FPs for $f_1$ and $f_2$ are the zeroes of the modified beta functions
$$
\tilde\beta_{f_i}=\frac{df_i}{d\tilde t}=\frac{1}{\lambda}\beta_{f_i}\,.
$$
They are just polynomials in $f_1$ and $f_2$.
The model has no real FP for $n=2$, but there are FPs for all $n>2$.
For $n=3,\ldots 8$ they are given in the fifth and sixth column in Table I.
One can then insert the FP values of $f_1$ and $f_2$ in $\beta_{\tilde g^2}$
and look for FP of $\tilde g^2$.
In each case there are two solutions, one at $\tilde g^2=0$, the other at
some nonzero value. These solutions are reported in the fourth column in Table I,
for $n=3,\ldots 8$.
The first solution describes the Gaussian FP (GFP), 
where all the couplings $\tilde g^2$, $\lambda$, $1/\ell_1$, $1/\ell_2$ are zero, 
the others non Gaussian FP's (NFP) where $\tilde g^2$ has finite limits instead.
Each FP can be approached only from specific directions in the space parametrized by 
$\lambda$, $\ell_1$, $\ell_2$, {\it i.e.} the ratios $f_1$ and $f_2$ take specific values.
For each NFP these values are unique, while for the GFP there may be several possible values: 
two if $n=3,4,5$ and four if $n=6,7,8$.

When one considers the linearized flow around any of the GFPs,
one finds as expected that the critical exponents, defined as minus the eigenvalues
of the matrix $\frac{\partial\beta_i}{\partial g_j}$, 
are -2,0,0,0, corresponding to the canonical dimensions of the couplings.
The critical exponents at the NGP are instead 2,0,0,0.
Thus the dimensionless couplings are marginal, and of the two FPs,
the trivial one is IR attractive and the nontrivial one UV attractive for $\tilde g$.
For $\lambda$ it is clear that the FP is UV attractive
(if we had chosen $\lambda<0$ it would be IR attractive). 
In order to establish the attractive or repulsive character of $f_1$ and $f_2$, 
one can look at the linearized flow in the variable $\tilde t$,
which is described by the $2\times2$ matrix 
$$
\frac{\partial\tilde\beta_{f_i}}{\partial f_j}\,.
$$
We define the ``critical exponents'' $\theta_{1,2}$ to be minus the eigenvalues
of this matrix. They are reported in the last two columns of table I, for $n=3,\ldots 8$.
It is important to realize that even for the GFP the eigenvectors of the stability matrix are
not the operators that appear in the action but mixings thereof.
We do not report the eigenvectors here.

\begin{table}
\begin{center}
\begin{tabular}{|l|l|l|l|l|l|l|l|l|}  \hline
$n$ & $\tilde g_*^{(III)}$\ & FP  &\ $\tilde g_*$\ & \ $f_{1*}$\ &\ $f_{2*}$\ &\ $\theta_1$\ &\ $\theta_2$\ \\
\hline
3& 8.886 & NFP1 & 6.626 & -0.693 & 0.453 &\ 0.094 & -0.0121 \\
3&       & NFP2 & 6.390 & -1.042 & 0.615 &\ 0.103 &\ 0.0119 \\
3&       & GFP1 & 0     & -0.693 & 0.453 &\ 0.094 & -0.0121 \\
3&       & GFP2 & 0     & -1.042 & 0.615 &\ 0.103 &\ 0.0119 \\
\hline
4& 7.255 & NFP1 & 5.877 & -0.479 & 0.398 &\ 0.105 & -0.0412 \\
4&       & NFP2 & 5.442 & -1.555 & 0.852 &\ 0.132 &\ 0.0392 \\
4&       & GFP1 & 0     & -0.479 & 0.398 &\ 0.105 & -0.0412 \\
4&       & GFP2 & 0     & -1.555 & 0.852 &\ 0.132 &\ 0.0392 \\
\hline
5& 6.283 & NFP1 & 5.310 & -0.400 & 0.400 &\ 0.118 & -0.0608 \\
5&       & NFP2 & 4.924 & -1.875 & 0.988 &\ 0.154 &\ 0.0567 \\
5&       & GFP1 & 0     & -0.400 & 0.400 &\ 0.118 & -0.0608 \\
5&       & GFP2 & 0     & -1.875 & 0.988 &\ 0.154 &\ 0.0567 \\
\hline
6& 5.620 & NFP1 & 4.883 & -0.350 & 0.408 &\ 0.131 & -0.0780 \\
6&       & NFP2 & 4.577 & -2.131 & 1.091 &\ 0.171 & -0.0717 \\
6&       & GFP1 & 0     & -0.350 & 0.408 &\ 0.131 & -0.0780 \\
6&       & GFP2 & 0     & -2.131 & 1.091 &\ 0.171 &\ 0.0717 \\
6&       & GFP3 & 0     & -0.814 & 1.369 & -0.161 & -0.0539 \\
6&       & GFP4 & 0     & -2.363 & 2.091 & -0.164 & -0.0617 \\
\hline
7& 5.130 & NFP1 & 4.548 & -0.314 & 0.417 &\ 0.143 & -0.0939 \\
7&       & NFP2 & 4.322 & -2.347 & 1.175 &\ 0.185 &\ 0.0851 \\
7&       & GFP1 & 0     & -0.314 & 0.417 &\ 0.143 & -0.0939 \\
7&       & GFP2 & 0     & -2.347 & 1.175 &\ 0.185 &\ 0.0851 \\
7&       & GFP3 & 0     & -2.790 & 2.130 & -0.181 & -0.0647 \\
7&       & GFP4 & 0     & -0.598 & 1.241 & -0.174 & -0.0716 \\
\hline
8& 4.750 & NFP1 & 4.274 & -0.286 & 0.424 &\ 0.156 & -0.1092 \\
8&       & NFP2 & 4.125 & -2.535 & 1.247 &\ 0.197 &\ 0.0976 \\
8&       & GFP1 & 0     & -0.286 & 0.424 &\ 0.156 & -0.1092 \\
8&       & GFP2 & 0     & -2.535 & 1.247 &\ 0.197 &\ 0.0976 \\
8&       & GFP3 & 0     & -2.790 & 2.131 & -0.180 &\ 0.1023 \\
8&       & GFP4 & 0     & -0.598 & 1.247 & -0.187 & -0.0872 \\
\hline
\end{tabular}
\end{center}
\caption{Gaussian and non-Gaussian fixed points of the $S^n$ model at one loop.
The first column gives the dimension $n$. The second column gives the
position of the NGFP in the two-derivative truncation, using a type III cutoff.
The rest of the table refers to the four-derivative truncation, also using a type III cutoff.
The third column gives the name of the FP. Columns 4,5,6 give the
position of the NGFP, columns 7,8 the critical exponents, as defined in the text. 
The coupling $\lambda$, not listed, goes to zero and is marginal in this approximation.}
\label{table1}
\end{table}

Beyond the values given in Table I, we have checked numerically the existence of the FP up to $n=200$.
For large $n$ one can study the theory analytically, to some extent.
There are four FPs for the system of the $f_i$'s, which are:
$f_1=0,\,f_2=1$ with critical exponents $\theta_1=6,\,\theta_2=12$,
$f_1=0,\,f_2=1/2$ with critical exponents $\theta_1=6,\,\theta_2=-12$,
$f_1=-6,\,f_2=5/2$ with critical exponents $\theta_1=-6,\,\theta_2=12$,
$f_1=-6,\,f_2=2$ with critical exponents $\theta_1=-6,\,\theta_2=-12$.
The numerical values at finite $n$ do indeed tend towards these limits for growing $n$.

\subsection{The chiral models}

The chiral model with $N=2$ is equivalent to the spherical model with $n=3$
(up to the redefinition of the couplings mentioned in the end of section IIID)
so we need not discuss this case further.
For ease of comparison we just report the properties of its nontrivial FPs
in the parametrization we used for the chiral models:
\begin{eqnarray}
NFP1: &&f_{1*}=-0.173\ ;\qquad f_{2*}=0.113\ ;\qquad \tilde g=13.25\nonumber \\
NFP2: &&f_{1*}=-0.261\ ;\qquad f_{2*}=0.154\ ;\qquad \tilde g=12.78\nonumber
\end{eqnarray}
The critical exponents don't depend on the definition of the couplings and therefore
are the same as in Table I;
they do however depend on the choice of RG parameter and they
differ from those given in \cite{hasenfratz} by a factor
$4\pi^2$, which is due to the definition of the parameter $x$ there.

In the case $N=3$ the system of the $f_i$'s has two FPs at
\begin{eqnarray}
FP1:&&\ \ f_{1*}=-0.154\ ;\qquad f_{2*}=0.050\ ;\qquad f_{3*}=0.085\ ;\nonumber\\
FP2:&&\ \ f_{1*}=-0.108\ ;\qquad f_{2*}=0.043\ ;\qquad f_{3*}=0.061\ .\nonumber
\end{eqnarray}
The attractivity properties in the space spanned by the $f_i$'s is given,
as in the spherical case, by studying the modified flow with parameter $\tilde t$.
The critical exponents at FP1 are: 
0.0303 with eigenvector (0.411, 0.630, 0.658);
0.0123 with eigenvector (0.515, -0.570, 0.640);
0.00289 with eigenvector (0.869, -0.148, -0.473),
whereas at FP2 they are:
0.0280  with eigenvector (0.366, 0.618, 0.695);
0.0108  with eigenvector (0.513, -0.575, 0.638) and
-0.00293 with eigenvector (0.887, -0.125, -0.445).
Therefore FP1 is attractive in all three directions,
while FP2 is attractive in two directions.
For each of these two FP's, the beta function of $\tilde g$ has two FP's:
the trivial FP, which has always critical exponents -2,
and a nontrivial FP, which is located at $\tilde g=11.17$ for NFP1
or $11.50$ for NFP2, and having critical exponent 2 in both cases.

We have found no FP's for $N>3$:
the system of equations $\tilde\beta_{f_i}=0$ for $i=1,2,3,4$ only has complex solutions.
To cover all of theory space we have checked this statement also
in the parametrization of the $\ell_i$ and in the parametrization of $u_i=1/\ell_i$.
This is true also in the large $N$ limit.
If we keep only the leading terms (of order $N^2$ for $f_1$ and $f_2$
and of order $N$ for $f_3$ and $f_4$), again the resulting polynomials
do not have any real zero.

\section{Discussion}

We have calculated the one loop beta functionals of the NLSM with values in any manifold,
in the presence of a very general class of four derivative terms.
We have then specialized our results to two infinite families of models: 
the $O(N)$ models, with values in spheres, and the chiral models
with values in the groups $SU(N)$.
Such calculations had been done before, but since the results are rather
complicated, it is useful to have independent verifications.
Our approach is calculationally very similar to \cite{bk},
but after correcting some small errors at the general level,
we find that the FP structure of the $O(N)$ models is completely different
from their findings.
On the other hand our results for the chiral models agree completely
with \cite{hasenfratz} for what concerns the dimensionless couplings, 
even though the calculation was done using very different techniques.
Since $SU(2)=S^3$, this provides a check also for our results for the spheres.

Since our aim is to establish asymptotic safety, or lack thereof,
it is important for us to have also the beta functions of the dimensionful
coupling $g$, which in the chiral models is the inverse of the pion decay constant.
This had not been considered at all in \cite{bk}, but it had been calculated
in \cite{hasenfratz} for the chiral models. Again we have agreement with the
result of \cite{hasenfratz}, up to a single factor 2 in one term; as discussed before, 
since this beta function is scheme dependent, we believe
that this is not an error on either side, but the result of the different way
in which the calculation was done.
This difference results in a shift of the FP value of $\tilde g$;
for example in the case of $SU(2)$ one would find $\tilde g=19.88$ instead of 13.25 for NFP1
and 18.39 instead of 12.78 for NFP2.
Such variations by a factor of order 2 are to be expected.

One of the motivations of this work was to use the NLSM as a toy model for gravity.
From this point of view we have a perfect correspondence of results.
If we use the $1/p^2$ propagator that comes from the two derivative term,
both theories are perturbatively unitary but nonrenormalizable;
if on the other hand we use the $1/p^4$ propagator that comes from
the four derivative terms both theories are renormalizable 
(see \cite{stelle} for gravity and \cite{slavnov} for the NLSM) but contain ghosts. 
In the latter case it had also been established (see \cite{julve,fradkin,avramidi,shapiro} 
for gravity and \cite{hasenfratz, bk} for the NLSM) that the four derivative terms,
whose couplings are dimensionless, are asymptotically free.
Actually the analogy works even in greater detail.
The coefficient of the square of the Weyl tensor (for gravity) 
and the square of $\Box\varphi^\alpha$ (for the NLSM)
have at one loop a beta function that is constant.
These coefficients diverge logarithmically in the UV,
so their inverses, which are the perturbative couplings,
are asymptotically free.
The coefficients of the other four derivative terms have more complicated
beta functions, but overall there is asymptotic freedom, provided the
Gaussian FP is approached from some special direction.
There have been many attempts to avoid the effects of the ghosts,
see \cite{julve,salam} for gravity and \cite{hasenfratz} for the NLSM.
In any case, the existence of the ghosts is only established at tree level.
Whether they exist in the full quantum theory is a deep dynamical question
whose answer is not known.

All these ``old'' works on higher derivatives theories concentrated on the
behavior of the couplings that multiply the four derivative terms;
much less attention, if any, was payed to the coefficient of the two derivative term, 
which has dimension of square of a mass:
the inverse of Newton's constant in gravity and
the square of the pion decay constant in the chiral NLSM.
In several papers this issue was ignored, or incorrect results were given,
because of the use of dimensional regularization.
The correct RG flow of these couplings is quadratic
in $k$, and is best seen when a momentum cutoff is used.
In \cite{codello1} this point was made for gravity and it was shown
that when this quadratic running is taken into account 
the beta function for Newton's constant (and for the cosmological constant) 
has, in addition to the Gaussian one, also a nontrivial FP.
In this paper we have found that the same is true in a large class of NLSM.
This is crucial for asymptotic safety.

It is somewhat gratifying to see that the FP does not always exist for all NLSM:
in particular we have seen that within the one loop approximation,
adding the higher derivative terms destroys the FP that is present in the
two derivative truncation for the sphere $S^2$ and for the chiral models with $N>3$.
If there was any doubt,
this shows that the existence of the FP is not ``built into the formalism'' 
but is a genuine property of the theory.
This is somewhat analogous to the situation when one adds minimally
coupled matter fields to gravity \cite{perini1}.

The next step will be to replace the one loop functional RG equation \eqref{onelooperge}
by its {\it exact} counterpart, which only differs in the replacement of
the bare action $S$ by $\Gamma_k$ in the r.h.s. \cite{wetterich}.
There are at least two good reasons to do this calculation.
One of the points of \cite{codello2} that needed further clarification was the value 
of the lowest critical exponent. In the two derivative truncation at one loop it was
always 2 at the nontrivial FP. 
Thus the critical exponent $\nu$ that governs the rate at which the correlation length 
diverges was given by
$$
\nu=-\frac{1}{\frac{d\beta}{dt}_*}=\frac{1}{\theta}=\frac{1}{2}\ ,
$$
which is the value of mean field theory.
Using the ``exact'' RG truncated at two derivatives gave $\nu=3/8$ for the $O(N)$ models,
independent of $N$.
One would like to understand what effect the higher derivative terms have on this exponent. 
Since here we restricted ourselves to one loop, we found again $\nu=1/2$,
so the calculations of this paper are of no use in this respect.
Another motivation comes from recent calculations in higher derivative gravity
\cite{bms} that go beyond one loop and find that the theory is not asymptotically free,
but rather all couplings reach nonzero values at the UV FP.
It would be interesting to see similar behaviour in (some) NLSM.

Concerning possible direct phenomenological applications of the NLSM,
regarded as an effective field theory,
it is interesting to ask what relation, if any, the UV properties
of the NLSM may have to the properties of the underlying fundamental theory.
Regarding the chiral NLSM as the low energy approximations of a QCD-like theory,
one may note that there is rough agreement between the range of existence
of the NLSM FP and the ``conformal window'' for the existence of an IR FP
in the case when the quarks are in the adjoint or in the symmetric tensor
representation \cite{sannino}.
One could get a better understanding of this issue if the beta functions of
the NLSM depended on the number of ``colors'' of the underlying theory,
which in the effective theory are reflected in the coefficient of the
Wess-Zumino-Witten term \cite{witten}.
The one loop beta function of the Wess-Zumino-Witten term is zero
\cite{braaten,bk}; 
this is consistent with the quantization of the coefficient $c$.
Unfortunately the beta functions of the remaining couplings are completely
independent of this coefficient, so the low energy theory seems to be
insensitive to this parameter.

Another possible application is to electroweak chiral perturbation theory \cite{ab}.
If the NLSM turned out to be asymptotically safe in the presence of gauge fields and
fermions, then one may envisage a higgsless standard model up to very high energies.
This will also require a separate investigation.
A related application of asymptotic safety to the standard model has been
discussed recently in \cite{gies}.

To summarize we believe that the NLSM are interesting theoretical laboratories
in which one may test various theoretical ideas, and they have also important
phenomenological applications.
The question whether some NLSM could be asymptotically safe seems to us
to be a particularly important one, and to deserve more attention.

\medskip
\centerline{Acknowledgements}
\noindent
We would like to thank J. Ambj\o rn and F. Sannino for 
discussions and S. Ketov for correspondence.

\goodbreak

\section{Appendix A}

In \cite{hasenfratz} the action for the chiral $SU(N)$ model is written in the form:
\begin{eqnarray}
\frac{1}{f^2}&&\!\!\!\!\!\!\int d^4x\,\Bigl(
c_0 \mathrm{Tr} L_\mu L^\mu
+\frac{1}{2}\mathrm{Tr}(\partial_\mu L^\mu\partial_\nu L^\nu+\partial_\mu L_\nu\partial^\mu L^\nu)
-\frac{1}{2}c_2\mathrm{Tr}(\partial_\mu L^\mu\partial_\nu L^\nu-\partial_\mu L_\nu\partial^\mu L^\nu)
\nonumber\\
&&\!\!\!\!\!\!
-\frac{1}{2}c_3\mathrm{Tr}(L_\mu L^\mu L_\nu L^\nu+L_\mu L_\nu L^\mu L^\nu)
-c_4\mathrm{Tr}(L_\mu L^\mu) \mathrm{Tr}(L_\mu L^\mu)
-c_5\mathrm{Tr}(L_\mu L^\mu) \mathrm{Tr}(L_\mu L^\mu)
\Bigr]\ .
\label{hasenfratzaction}
\end{eqnarray}
where $L_\mu=U^{-1}\partial_\mu U$.
We want to translate this action into the form \eqref{actionsun}.
Deriving the equation $L_\mu=\partial_\mu\varphi^\alpha L^a_\alpha (-iT_a)$
we obtain
$$
\partial_\mu L_\nu=-i T_a(\nabla_\mu\partial_\nu\varphi^\alpha L_\alpha^a
-\partial_\mu\varphi^\alpha\partial_\nu\varphi^\beta \nabla_\alpha L^a_\beta)
$$
The antisymmetric part of this equation is
$$
\partial_\mu L_\nu-\partial_\nu L_\mu=-[L_\mu,L_\nu]
$$
whereas using Killing's equation, the symmetric part is
$$
\partial_{(\mu} L_{\nu)}=-i T_a\nabla_\mu\partial_\nu\varphi^\alpha L_\alpha^a
$$
The terms of \eqref{hasenfratzaction} have the following translation into 
our
tensorial language:
\begin{eqnarray*}
\int d^4x\,\mathrm{Tr} L_\mu L^\mu&=&-\frac{1}{2}
\int d^4x\,\partial_\mu\varphi^\alpha\partial^\mu\varphi^\beta h_{\alpha\beta}
\\
\int d^4x\,\mathrm{Tr}\partial_\mu L^\mu\partial_\nu L^\nu&=&
-\frac{1}{2}\int d^4x\,
\Box\varphi^\alpha \Box\varphi^\beta h_{\alpha\beta}
\\
\int d^4x\,\mathrm{Tr}\partial_\mu L_\nu\partial^\mu L^\nu&=&
\!-\frac{1}{2}\!\int d^4x\!\left(
\nabla^\mu\partial^\nu\varphi^\alpha \nabla_\mu\partial_\nu\varphi^\beta h_{\alpha\beta}
+\frac{1}{4}
\partial_\mu\varphi^\alpha\partial^\mu\varphi^\beta
\partial_\nu\varphi^\gamma\partial^\nu\varphi^\delta
T^{(3)}_{\alpha\beta\gamma\delta}\right)
\\
\int d^4x\,\mathrm{Tr}L_\mu L^\mu L_\nu L^\nu&=&
\int d^4x\,
\partial_\mu\varphi^\alpha\partial^\mu\varphi^\beta
\partial_\nu\varphi^\gamma\partial^\nu\varphi^\delta
\left(\frac{1}{4N}T^{(2)}_{\alpha\beta\gamma\delta}
+\frac{1}{8}T^{(5)}_{\alpha\beta\gamma\delta}\right)
\\
\int d^4x\,\mathrm{Tr}L_\mu L_\nu L^\mu L^\nu&=&
\int d^4x\,
\partial_\mu\varphi^\alpha\partial^\mu\varphi^\beta
\partial_\nu\varphi^\gamma\partial^\nu\varphi^\delta
\left(\frac{1}{4N}T^{(1)}_{\alpha\beta\gamma\delta}
-\frac{1}{8}T^{(3)}_{\alpha\beta\gamma\delta}
+\frac{1}{8}T^{(4)}_{\alpha\beta\gamma\delta}\right)
\\
\int d^4x\,\mathrm{Tr}(L_\mu L^\mu) \mathrm{Tr}(L_\nu L^\nu)&=&  
\frac{1}{4}\int d^4x\,
\partial_\mu\varphi^\alpha\partial^\mu\varphi^\beta
\partial_\nu\varphi^\gamma\partial^\nu\varphi^\delta
T^{(2)}_{\alpha\beta\gamma\delta}
\\
\int d^4x\,\mathrm{Tr}(L_\mu L_\nu) \mathrm{Tr}(L^\mu L^\nu)&=&
\frac{1}{4}\int d^4x\,
\partial_\mu\varphi^\alpha\partial^\mu\varphi^\beta
\partial_\nu\varphi^\gamma\partial^\nu\varphi^\delta
T^{(1)}_{\alpha\beta\gamma\delta}
\end{eqnarray*}
where $\Box\varphi^\alpha=\nabla^\mu\partial_\mu\varphi^\alpha$.
One can further manipulate the third term integrating by parts and
commuting covariant derivatives. One finds
$$
\int d^4x\,
\nabla^\mu\partial^\nu\varphi^\alpha \nabla_\mu\partial_\nu\varphi^\beta h_{\alpha\beta}
=\int d^4x\,\left(
\Box\varphi^\alpha \Box\varphi^\beta h_{\alpha\beta}
+\partial_\mu\varphi^\alpha\partial^\mu\varphi^\beta
\partial_\nu\varphi^\gamma\partial^\nu\varphi^\delta
R_{\alpha\beta\gamma\delta}\right)
$$
and using \eqref{curvaturesun} one can further substitute the Riemann tensor by $T^{(3)}$.
In the fourth term one can eliminate $T^{(5)}$.

One has to note that Hasenfratz's action has to be compared to {\it minus} 
our action. This is because it appears with the positive sign in the
exponent of the functional integral (this is consistent with the fact
that the $(\Box\varphi)^2$ term has a negative coefficient in \eqref{hasenfratzaction}).
It is then straightforward to calculate the following relations between the 
couplings used in \cite{hasenfratz} and our couplings:
$$
g^2=\frac{f^2}{c_0}\ ;\qquad 
\lambda=f^2\ ;\qquad
f_1=\frac{c_3}{2N}+\frac{c_5}{2}\ ;\qquad
f_2=\frac{c_4}{2}\ ;\qquad
f_3=\frac{1+c_2}{4}\ ;\qquad
f_4=\frac{c_3}{4}\ ;\qquad
$$

With these relations, one can translate his beta functions and one finds that
they agree with those given in section IIID, with a single exception: 
the term proportional to $\tilde g^4$ and containing no $f_i$ in $\beta_{\tilde g^2}$.
We observe that the two polynomials in the $c$'s
in equation (39) in \cite{hasenfratz} are the same, up to an overall factor 2.
As a consequence, when one extracts the beta function of $c_0/f^2=1/g^2$ and
rewrites it in terms of the $f_i$'s, the coefficients of $\tilde g^4$ and $\tilde g^2\lambda$
are exactly the same. This differs from the beta function given in (\ref{bsung}),
where the two coefficients differ in the last term.
We believe that this difference can be attributed to the different cutoff scheme.

\end{document}